\documentclass[twocolumn]{aastex701}
\journalinfo{}
\makeatletter
\def\frontmatter@title@above{}%
\makeatother

\usepackage{natbib}
\setcitestyle{super}

\usepackage[version=3]{mhchem} 
\usepackage{comment}
\usepackage{amssymb}
\usepackage{mathrsfs}
\usepackage{orcidlink} 

\renewcommand{\thefootnote}{\alph{footnote}}

\newcommand{\be}{\begin{equation}}
\newcommand{\ee}{\end{equation}}
\newcommand{\ud}{{\rm d}}
\newcommand{\mk}{\mathcal{K}}

\makeatletter
\renewcommand{\doi}[1]{\@ifnextchar.{\@gobble}{}}
\renewcommand{\url}[1]{\@ifnextchar.{\@gobble}{}}
\makeatother

\begin{document}

\title{Excitation of molecular hydrogen by cosmic-ray protons}

\author{Marco Padovani\,\orcidlink{0000-0003-2303-0096}}
\affiliation{INAF--Osservatorio Astrofisico di Arcetri, Largo E. Fermi 5, 50125 Firenze, Italy}
\email[show]{marco.padovani@inaf.it}  

\author{Daniele Galli\,\orcidlink{0000-0001-7706-6049}} 
\affiliation{INAF--Osservatorio Astrofisico di Arcetri, Largo E. Fermi 5, 50125 Firenze, Italy}
\email{a@a.a}  

\author{Corey~T.~Plowman\,\orcidlink{0000-0001-8224-7143}}
\affiliation{Department of Physics, Curtin University, Perth, WA 6845, Australia}
\email{a@a.a}  

\author{Liam~H.~Scarlett\,\orcidlink{0000-0002-9900-9712}}
\affiliation{Department of Physics, Curtin University, Perth, WA 6845, Australia}
\email{a@a.a}  

\author{Mark~C.~Zammit\,\orcidlink{0000-0003-0473-379X}}
\affiliation{Theoretical Division, Los Alamos National Laboratory, Los Alamos, New Mexico 87545, USA}
\email{a@a.a}  

\author{Igor~Bray\,\orcidlink{0000-0001-7554-8044}}
\affiliation{Department of Physics, Curtin University, Perth, WA 6845, Australia}
\email{a@a.a}  

\author{Dmitry~V.~Fursa\,\orcidlink{0000-0002-3951-9016}}
\affiliation{Department of Physics, Curtin University, Perth, WA 6845, Australia}
\email{a@a.a}

\begin{abstract}

Low-energy cosmic rays ($E\lesssim 1$~GeV) are responsible for the ionisation and heating of molecular clouds. While the role of supra-thermal electrons produced in the ionisation process in inducing excitation of the ambient gas (mostly molecular hydrogen) has been studied in detail, the role of primary cosmic-ray nuclei (protons and heavier nuclei) has been generally neglected.
Here, we introduce, for the first time, 
cross sections for proton impact on \ce{H2}, calculated using the semi-classical implementation of the molecular convergent close-coupling method. 
Our findings show that proton-induced \ce{H2} excitation is comparable in magnitude to that caused by electrons. 
We discuss the possible implications on the estimate of the cosmic-ray ionisation rate from observations in the near-infrared domain and 
on the cosmic-ray-induced \ce{H2} ultraviolet luminescence.
We also derive a new approximated analytical parameterisation of the spectrum of secondary electrons that can be easily incorporated in numerical codes.

\end{abstract}


\section{Introduction}

Cosmic rays with energies between a few MeV and a few GeV are 
the dominant source of ionisation of the dense, cold regions of molecular clouds shielded against UV and X-ray photons 
(visual extinctions $A_V\gtrsim3-4$~mag). 
The primary ionisation of \ce{H2} by low-energy cosmic rays produces \ce{H2+} that rapidly converts to
\ce{H3+}, initiating the ion-molecule reaction chains that control chemical enrichment in cold molecular clouds ($T\sim10-50$~K)\cite{HerbstKlemperer1973,HerbstMillar2008}.
An important consequence of the ionisation of ambient molecular hydrogen (and other species) is the generation of a flux of supra-thermal secondary electrons
with average energies in the range of $30-50$~eV\cite{CravensDalgarno1978,GredelDalgarno1995,Ivlev+2021,Padovani+2022,Padovani+2024},
which represents a major contribution to the excitation of 
rovibrational levels of \ce{H2}, both in the ground and in excited electronic states. 
The subsequent radiative decay of collisionally excited \ce{H2} results in the emission of photons in the near-infrared (NIR) and mid-infrared (MIR) domain, for rovibrational and pure rotational transitions in the ground state, respectively, and at ultraviolet (UV) wavelengths, for electronic 
transitions\cite{PrasadTarafdar1983,Gredel+1989}. 
These UV photons generated by cosmic rays, together with the excitation and ionisation caused by cosmic rays and secondary electrons, pave the way for both gas-phase photochemistry and non-thermal processing of icy mantles. The latter processes, collectively known as radiation chemistry, have been incorporated into astrochemical networks, quantifying how cosmic-ray irradiation produces suprathermal radicals and ions in ices and how these species feed reaction channels otherwise inaccessible at $\sim 10$~K\cite{ShingledeckerHerbst2018}.
When radiation chemistry and grain surface processes are included in a self-consistent manner, models show that low-energy cosmic rays directly increase the production of complex organic compounds in cold cores, both by generating reactive ions in the gas phase and by producing species in the solid phase that drive bulk and surface chemistry\cite {Hasegawa1992,ShingledeckerTennis2018,Shingledecker2019,Pilling2023,daSilveira2024,Pilling2024,Silva2025}.

A case in point has been the prediction of the intensity of NIR emission due to the flux of secondary electrons generated by cosmic rays \cite{Bialy2020,Padovani+2022,Bialy+2022,Gaches+2022}, which has recently been confirmed by JWST observations\cite{Bialy+2025,Neufeld+2025}. Achieving this result has also been possible thanks to the availability of accurate \ce{H2} rovibrational cross sections calculated with the molecular convergent close-coupling (MCCC) method \cite{Padovani+2022,Scarlett+2023,Scarlett+2021a,ZFSB17review}.
The MCCC method is a fully quantum-mechanical technique with a 
highly-efficient computational implementation specifically optimised for
rovibrationally-resolved studies of electron and positron collisions 
with simple molecules of astrophysical and fusion interest. 
So far it has been applied to the 
\ce{H2+}, \ce{H2}, \ce{HeH+}, \ce{LiH}, \ce{Li2}, and \ce{H3+}
molecules\cite{Zammit2014_H2+,Scarlett2017_H2+KER,Zammit+2017,
Zammit2017_pos-H2,Scarlett+2021a,Scarlett22HeH+,Umer2025LiH,
Umer2025Li2,Horton2025,Horton2025b}, including various isotopologues 
(see the online database\setcounter{footnote}{0}\footnote{\href{https://mccc-db.org}{https://mccc-db.org}} for up-to-date results and references).
In addition, based on fundamental work\cite{Roberge1983,PrasadTarafdar1983,Sternberg1987,Gredel+1987,Gredel+1989}, the calculation of these new rotationally-resolved excitation cross sections for excited electronic states\cite{Scarlett+2021a,Scarlett+2023} also allowed the calculation
of the UV \ce{H2} luminescence induced by cosmic rays with unprecedented accuracy\cite{Padovani+2024}.

\section{Methods}

The results obtained so far have suffered from the lack of data on the proton-impact excitation of \ce{H2}. 
Experimental estimates for the excitation of selected vibrational bands of the Lyman system have been reported for proton energies in 
the $20-130$~keV range and above 150~keV\cite{Dahlberg1968,EdwardsThomas1968}, but the accuracy of these results has been questioned\cite{Thomas1972}. 
The majority of prior research has been directed towards studies of ionisation and electron capture, with only
a few previous calculations of the elastic and total excitation cross sections, and none for individual
excitation channels.
A common approximation, based on Bethe-Born theory, is to assume that the proton-impact excitation cross section is equal to
the cross section of an electron with the same velocity\cite{Cecchi-PestelliniAiello1992,Padovani+2024}, namely
\be
\sigma^{\rm exc}_{p}(E_p) \simeq \sigma^{\rm exc}_{e}\left(E_e=\frac{m_e}{m_p}E_p\right)\,,
\label{eq:equivelocity}
\ee
where $m_e$ and $m_p$ are the electron and proton mass, respectively, and $E_e$ and $E_p$ their corresponding energies.

To remedy the lack of data for proton-impact excitation of \ce{H2}, and examine the accuracy of the equal-velocity approximation, the MCCC method has recently been extended to study proton 
scattering\cite{Plowman2025_H2}. 
In this article, we apply newly available vibrationally resolved cross sections 
to investigate the contribution of protons to the excitation of 
electronic states $2p\sigma\,^1\Sigma_u^+$
and $2p\pi\,^1\Pi_u$ (hereafter $B$ and $C$, respectively). 
We show that protons play a role comparable to that of electrons and possibly even greater. 

\subsection{The semi-classical MCCC method}
\label{sect:xsecs}

Recently, a semi-classical implementation of the MCCC method (SC-MCCC) has been developed to allow the 
study of heavy-particle collisions with diatomic molecules, with a focus on producing 
rovibrationally resolved cross sections for both electronically elastic and inelastic collisions.
In the SC-MCCC method, the \ce{H2} target is treated in the same way as in the MCCC approach to electron collisions\cite{Scarlett+2021a,Scarlett+2023},
while the projectile is modelled as a classical particle.
Cross sections for inelastic processes are 
most significant at energies above 1~keV, where this semi-classical model is known to be very accurate~\cite{BM92}.
Apart from the straight-line treatment of the projectile motion, the SC-MCCC approach is
an ab-initio quantum-mechanical technique with the same level of accuracy as the 
MCCC method for electron scattering. 
The non-relativistic formulation is adopted for both the target and projectile description as \ce{H2} is a 
light molecule and the incident proton energies are not large. 
For proton scattering on the ground electronic state of \ce{H2} that is a singlet spin state, 
only singlet-state excitations are possible with excitations of $B$ and $C$ states being the 
lowest dipole-allowed transitions with the largest cross sections.
The uncertainty on the cross sections is estimated at the level of about 10\%.

The new implementation was first applied to the 
proton-He collision system\cite{Plowman2025_He} (by treating helium as an \ce{H2} molecule 
with an internuclear separation of zero), and yielded excellent agreement with the available 
experimental data. Following this validation, the method has been applied to proton collisions with
\ce{H2}\cite{Plowman2025_H2}, producing vibrationally resolved cross sections 
over an energy range spanning several orders of magnitude above 1~keV
for excitation of 11 low-lying electronic states, as well as elastic scattering, electron loss, and dissociation.
We checked that a linear extrapolation of the cross sections between the threshold and 1~keV in semi-log space 
resulted in virtually no change in the calculated excitation rates compared to assuming a zero cross section below 1 keV.

Aside from the contribution of a large set of previously unavailable data, one
of the key outcomes of the application of the SC-MCCC method to proton-\ce{H2} collisions\cite{Plowman2025_H2}
is the finding that the equal-velocity scaling approximation (Eq.~\ref{eq:equivelocity}) is much less accurate 
than previously thought. In this paper, 
we see that the implications of this finding are significant for the calculation of the excitation rates.
The approximation is only able to reproduce the SC-MCCC results at energies above 
$300-1000$~keV (depending on the transition), however the cross sections peak 
at around $100$~keV. A significant issue is that Eq.~(\ref{eq:equivelocity}) unphysically
scales the threshold energies from the $\sim 10$~eV region (for electronic excitation)
into the $\sim 20$~keV region, and hence the approximation returns a cross section of zero
over a substantial portion of the energy range where the true cross sections are in fact
reasonably large. This effect can be corrected by 
scaling the electron-impact cross sections to match the proton velocity after 
the collision, which maintains the correct excitation thresholds while still producing
good agreement with the SC-MCCC calculations in the high-energy limit\cite{Plowman2025_H2}. For systems where
proton-impact cross sections are not available, we suggest that the alternative equal-velocity scaling relation\cite{Plowman2025_H2} should be used in place of Eq.~(\ref{eq:equivelocity}). 
For the remainder of the present work, we no longer rely on the equal-velocity approximation, and instead use the accurate SC-MCCC data\footnote{The full set of proton-\ce{H2} cross section data used in this work will be openly available on the MCCC database (\href{https://mccc-db.org}{https://mccc-db.org}).}.

\subsection{Proton-impact excitation of \ce{H2}}
\label{sect:updateanalysis}

The ionisation and excitation rates of molecular hydrogen are given by
\begin{eqnarray}\label{eq:zion}
\zeta^{\rm ion}_{{\rm H}_2} &=& 4\pi \int_{I^{\rm ion}_{{\rm H}_2}}^\infty j_p(E) \sigma^{\rm ion}_{p,{\rm H}_2}(E)\,\ud E \\\nonumber
&&+4\pi \int_{I^{\rm ion}_{{\rm H}_2}}^\infty j_{\rm sec,\ce{H2}}(\varepsilon)\sigma^{\rm ion}_{e,{\rm H}_2}(\varepsilon)\,\ud\varepsilon \equiv
\zeta^{\rm ion}_{p,{\rm H}_2} + \zeta^{\rm ion}_{{\rm sec},{\rm H}_2}
\end{eqnarray}
and 
\begin{eqnarray}\label{eq:zexc}
\zeta^{\rm exc}_{{\rm H}_2} &=& 4\pi \int_{I^{\rm exc}_{{\rm H}_2}}^\infty j_p(E) \sigma^{\rm exc}_{p,{\rm H}_2}(E)\, \ud E \\\nonumber
&&+4\pi \int_{I^{\rm exc}_{{\rm H}_2}}^\infty j_{\rm sec,\ce{H2}}(\varepsilon)\sigma^{\rm exc}_{e,{\rm H}_2}(\varepsilon)\,\ud\varepsilon \equiv 
\zeta^{\rm exc}_{p,{\rm H}_2} + \zeta^{\rm exc}_{{\rm sec},{\rm H}_2}\,,
\end{eqnarray}
respectively.
Here, $E$ is the energy of cosmic-ray protons, $\varepsilon$ the energy of secondary electrons, $j_p(E)$ and $j_{{\rm sec},\ce{H2}}(\varepsilon)$ are respectively the proton and secondary electron specific intensity
in units of particles per unit energy, area, time, and solid angle,
$\sigma^{\rm ion}_{k,\ce{H2}}$ and $\sigma^{\rm exc}_{k,\ce{H2}}$ are the ionisation and excitation cross sections, respectively,
with the subscript $k=p,e$ denoting the incident particle (proton or electron),
$I^{\rm ion}_{{\rm H}_2}=15.44$~eV is the ionisation energy threshold and
$I^{\rm exc}_{{\rm H}_2}$ is the excitation energy threshold, which is transition dependent (from less than one eV
for rovibrational transitions within the ground electronic state to tens of eV for transitions to the excited electronic states). 

In Eq.~(\ref{eq:zion}), the ionisation rate $\zeta^{\rm ion}_{{\rm H}_2}$ is dominated by primary cosmic-ray nuclei,
plus a contribution of about 70\% from secondary electrons\cite{DalgarnoGriffing1958,Ivlev+2021}. 
On the other hand, 
the excitation rate of \ce{H2} in Eq.~(\ref{eq:zexc})
has generally been computed by including only the contribution of secondary electrons.

\section{Results}\label{sect:results}
\begin{figure}[!h]
\includegraphics[width=0.5\textwidth]{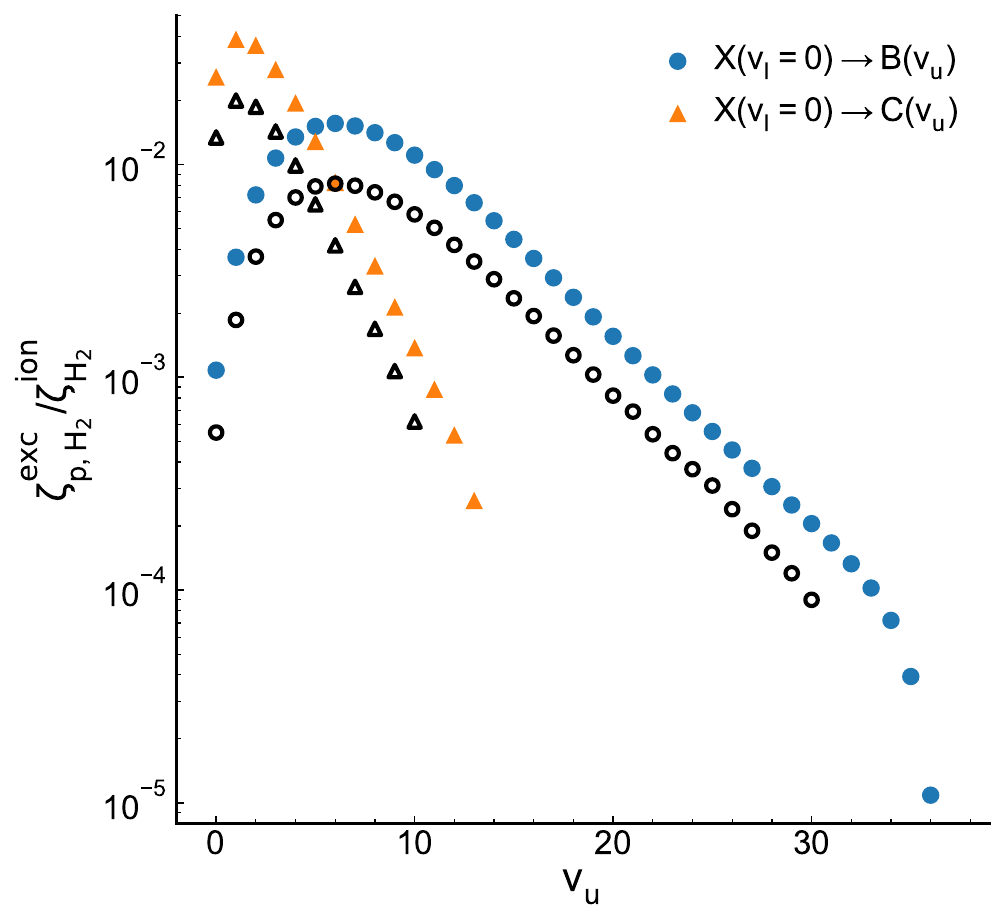}
\caption{Ratio between proton-impact \ce{H2} excitation rates ($\zeta^{\rm exc}_{p,{\rm H}_2}$) 
summed over the initial and final rotational states and the cosmic-ray ionisation rate ($\zeta^{\rm ion}_{{\rm H}_2}$)
as a function of the upper vibrational level, $v_u$. 
Results are shown for transitions from the lowest vibrational level of the ground state
$X(v_l=0)$ to the excited electronic states $B$ (solid blue circles) and $C$ (solid orange triangles).
Empty black symbols show a subset of previous estimates\cite{Cecchi-PestelliniAiello1992}.}
\label{fig:comparison_CP92_protons}
\end{figure}

Thanks to the new SC-MCCC data \cite{Plowman2025_H2}, it is now possible to assess more accurately the contribution of protons to the excitation of \ce{H2}.
First, we compare the ratios $\zeta^{\rm exc}_{p,{\rm H}_2}/\zeta^{\rm ion}_{{\rm H}_2}$ with those previously calculated\cite{Cecchi-PestelliniAiello1992} for non-rotationally resolved transitions from the lowest vibrational level, $v_l=0$, of the ground state $1s\sigma\,^1\Sigma^+_g$ (hereafter $X$ state) to the vibrational levels, $v_u$, of 
the excited electronic states $B$ and $C$. 
We note that this ratio is independent 
of the assumed interstellar cosmic-ray proton spectrum and 
of the \ce{H2} column density (see Sect.~\ref{sect:jsec}), so this ratio only reflects the relative magnitude of the underlying cross sections.
As shown in Fig.~\ref{fig:comparison_CP92_protons}, we find values of $\zeta^{\rm exc}_{p,{\rm H}_2}/\zeta^{\rm ion}_{{\rm H}_2}$
larger by $80-130$\% with respect to previous estimates. 
This result already points to potential implications on the role of protons.
It is worth mentioning that there is some uncertainty 
in how to define the dissociation limit. The energies for the vibrational levels 37, 38, and 39
of the excited electronic state $B$
are very close to the dissociation limit, 
reaching the limits of expected numerical accuracy in the structure calculations, which are based on the 
Born-Oppenheimer approximation.

\begin{figure*}[!h]
  \centering
  \begin{minipage}{0.48\textwidth}
    \includegraphics[width=\linewidth]{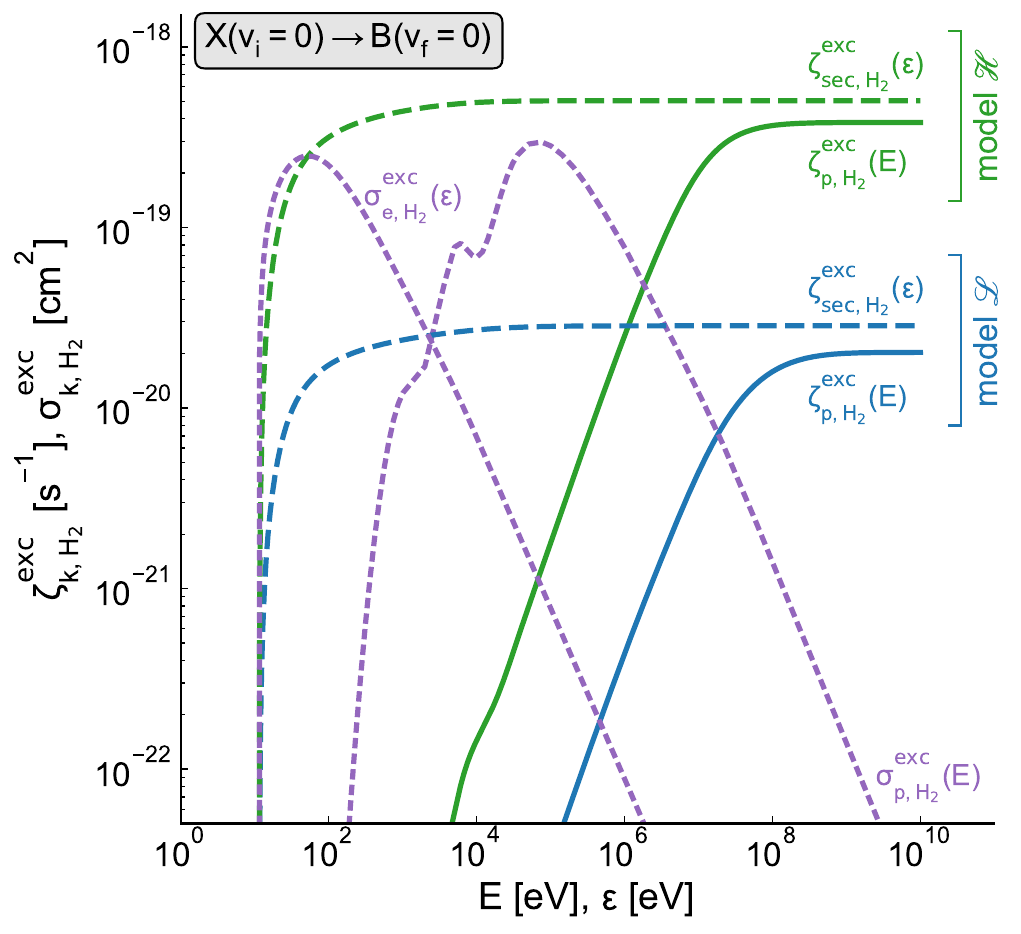}
  \end{minipage}
  \hfill
  \begin{minipage}{0.48\textwidth}
    \includegraphics[width=\linewidth]{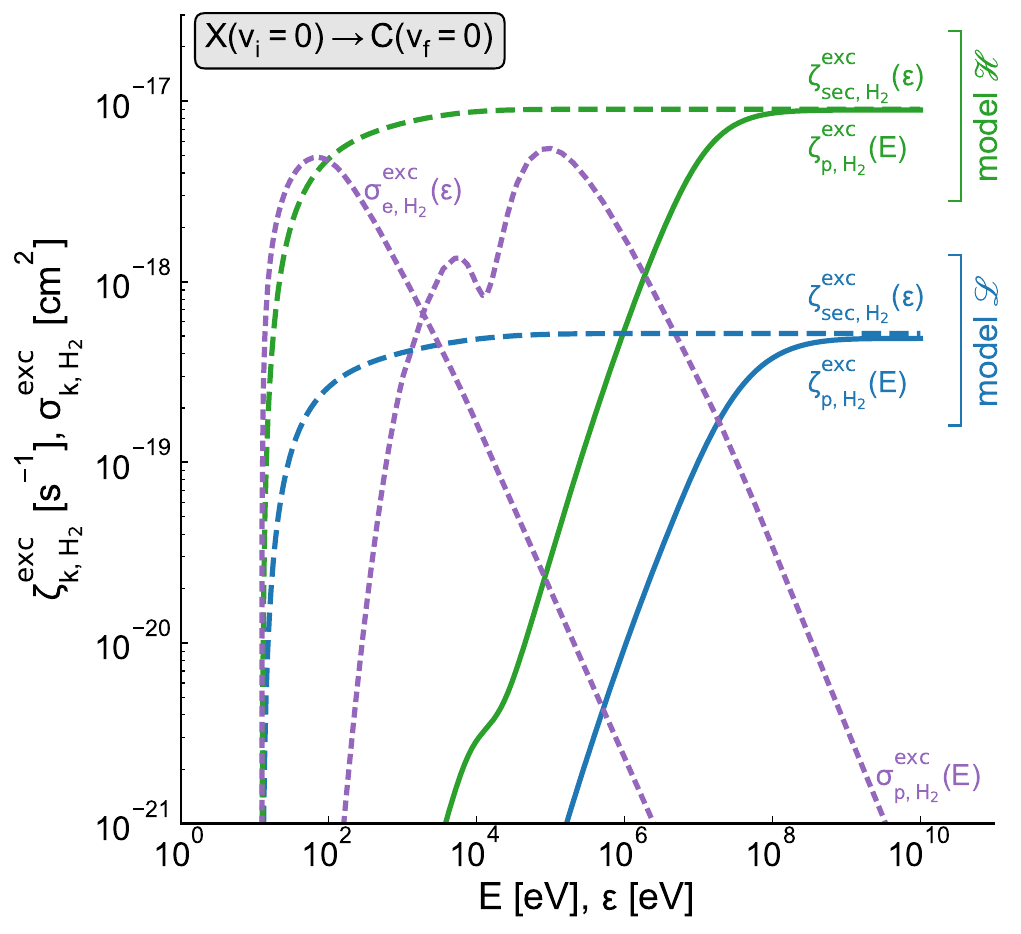}
  \end{minipage}
  \caption{Cumulative integral of the excitation rates of \ce{H2} by cosmic-ray protons and secondary electrons, $\zeta^{\rm exc}_{p,\ce{H2}}(E)$ and $\zeta^{\rm exc}_{{\rm sec},\ce{H2}}(\varepsilon)$,  
  and the corresponding excitation cross sections ($\sigma^{\rm exc}_{p,\ce{H2}}$ and $\sigma^{\rm exc}_{e,\ce{H2}}$, 
  purple short-dashed lines) for model $\mathscr{L}$
  and $\mathscr{H}$ (blue and green lines, respectively) computed at $N(\ce{H2})=10^{22}$~cm$^{-2}$.}
  \label{fig:zexc_cumulative}
\end{figure*}

In order to quantify how much protons affect the excitation rate compared to electrons, 
we calculate the contributions $\zeta^{\rm exc}_{p,{\rm H}_2}$ and $\zeta^{\rm exc}_{{\rm sec},{\rm H}_2}$ separately 
for all transitions $X(v_l=0)\rightarrow B(v_u)$ and $X(v_l=0)\rightarrow C(v_u)$.
For explanatory purposes, Fig.~\ref{fig:zexc_cumulative} shows excitation rates to the $v_u=0$ level of $B$ and $C$ states, for
two extreme models of the cosmic-ray proton flux: 
model $\mathscr{L}$, which reproduces the data from the Voyager probes\cite{Cummings+2016,Stone+2019}, 
and model $\mathscr{H}$, with an enhanced flux of sub-MeV protons (introduced to explain observational measurements of 
$\zeta^{\rm ion}_{{\rm H}_2}$ higher than expected from a Voyager-type interstellar spectrum\cite{Padovani+2018a}). 
Both spectra are propagated to a column density of molecular hydrogen 
of $N(\ce{H2})=10^{22}$~cm$^{-2}$, typical of molecular cloud
cores, using the continuous slowing-down approximation\cite{Takayanagi1973,Padovani+2009}. 
The cumulative integrals of the excitation rate shown in Fig.~\ref{fig:zexc_cumulative} highlight a 
fundamental difference between electrons and protons: while $\zeta^{\rm exc}_{{\rm sec},{\rm H}_2}$ converges rapidly, reaching 95\% of its final value at energies $\varepsilon$ between 5 and 20 keV, 
$\zeta^{\rm exc}_{p,{\rm H}_2}$ reaches 95\% of its final value at energies $E$ between $80$ and 400 MeV (for models $\mathscr{H}$ and $\mathscr{L}$, respectively).
This occurs because secondary electrons have a steep spectrum for
$\varepsilon\lesssim 100$~eV\cite{Ivlev+2021,Padovani+2022}, 
and because $\sigma^{\rm exc}_{e,{\rm H}_2}(\varepsilon)$ for transitions to $B$ and $C$ states peaks at $\varepsilon \approx 60-70$~eV.
On the other hand, the propagated proton spectra are rather flat\cite{Padovani+2022} and $\sigma^{\rm exc}_{p,{\rm H}_2}(E)$ peaks at $E\approx 100$~keV,
with a maximum contribution to excitation at
$E\approx 10-30$~MeV.

Tables~\ref{tab:ratios_B} and~\ref{tab:ratios_C} summarise the ratios $\zeta^{\rm exc}_{{\rm H}_2}/\zeta^{\rm ion}_{{\rm H}_2}$ 
and $\zeta^{\rm exc}_{{\rm sec},{\rm H}_2}/\zeta^{\rm exc}_{p,{\rm H}_2}$ for all the vibrational transitions of the new cross section data set.
For the calculation, we choose an interstellar cosmic-ray proton model at 
$N(\ce{H2})=10^{22}$~cm$^{-2}$ and solve the balance equation for the exact calculation of the secondary electron flux\cite{Ivlev+2021}.
It is evident that the contribution of cosmic-ray protons to the excitation of \ce{H2} cannot be neglected, since $\zeta^{\rm exc}_{{\rm sec},{\rm H}_2}/\zeta^{\rm exc}_{p,{\rm H}_2}$ varies between 0.871 to 1.40. 
This means that protons 
contribute at least as much as secondary electrons to excitation of \ce{H2} in molecular clouds. The consequences of this result are addressed in Sect.~\ref{sect:conclusions}.

\begin{table*}[htbp]
\centering
\caption{Ratio of the total excitation rate to the total ionisation rate ($\zeta^{\rm exc}_{\ce{H2}}/\zeta^{\rm ion}_{\ce{H2}}$) 
and of the excitation rate by secondary electron impact to the excitation rate by proton impact ($\zeta^{\rm exc}_{{\rm sec},\ce{H2}}/\zeta^{\rm exc}_{p,\ce{H2}}$)
of molecular hydrogen. Values are computed from the exact solution of 
the balance equation\cite{Ivlev+2021} for the spectrum of secondary electrons, and from the SS approximation (Eq.~\ref{eq:jsec_essa}), 
with their relative differences, for vibrational transitions $X(v_l=0)\rightarrow B(v_u)$.}
\label{tab:ratios_B}
\begin{tabular}{ccccccc}
\hline
$v_u$ & \multicolumn{2}{c}{$\zeta^{\text{exc}}_{{\rm H}_2}/\zeta^{\text{ion}}_{{\rm H}_2}$} & Rel. diff. (\%) & \multicolumn{2}{c}{$\zeta^{\rm exc}_{{\rm sec},\ce{H2}}/\zeta^{\rm exc}_{p,\ce{H2}}$} & Rel. diff. (\%) \\
\cline{2-3}\cline{5-6}
& exact & SS approx. & & exact & SS approx. & \\
\hline\hline
0 & $2.60(-3)$ & $1.74(-3)$ & $-33.1$ & $1.40$ & $1.22$ & $-13.0$ \\
1 & $8.61(-3)$ & $5.79(-3)$ & $-32.8$ & $1.35$ & $1.18$ & $-12.7$ \\
2 & $1.66(-2)$ & $1.12(-2)$ & $-32.4$ & $1.31$ & $1.14$ & $-12.4$ \\
3 & $2.44(-2)$ & $1.66(-2)$ & $-32.1$ & $1.28$ & $1.12$ & $-12.2$ \\
4 & $3.03(-2)$ & $2.07(-2)$ & $-31.8$ & $1.25$ & $1.10$ & $-11.9$ \\
5 & $3.36(-2)$ & $2.30(-2)$ & $-31.5$ & $1.22$ & $1.08$ & $-11.7$ \\
6 & $3.43(-2)$ & $2.36(-2)$ & $-31.3$ & $1.20$ & $1.06$ & $-11.5$ \\
7 & $3.30(-2)$ & $2.28(-2)$ & $-31.0$ & $1.18$ & $1.05$ & $-11.2$ \\
8 & $3.04(-2)$ & $2.11(-2)$ & $-30.7$ & $1.16$ & $1.03$ & $-11.0$ \\
9 & $2.70(-2)$ & $1.88(-2)$ & $-30.5$ & $1.14$ & $1.01$ & $-10.8$ \\
10 & $2.34(-2)$ & $1.63(-2)$ & $-30.2$ & $1.12$ & $1.00$ & $-10.6$ \\
11 & $1.99(-2)$ & $1.39(-2)$ & $-30.0$ & $1.10$ & $9.87(-1)$ & $-10.4$ \\
12 & $1.66(-2)$ & $1.17(-2)$ & $-29.8$ & $1.09$ & $9.76(-1)$ & $-10.2$ \\
13 & $1.37(-2)$ & $9.65(-3)$ & $-29.6$ & $1.07$ & $9.66(-1)$ & $-10.0$ \\
14 & $1.12(-2)$ & $7.92(-3)$ & $-29.4$ & $1.06$ & $9.56(-1)$ & $-9.8$ \\
15 & $9.11(-3)$ & $6.45(-3)$ & $-29.2$ & $1.05$ & $9.47(-1)$ & $-9.7$ \\
16 & $7.37(-3)$ & $5.23(-3)$ & $-29.0$ & $1.04$ & $9.39(-1)$ & $-9.5$ \\
17 & $5.94(-3)$ & $4.23(-3)$ & $-28.8$ & $1.03$ & $9.30(-1)$ & $-9.3$ \\
18 & $4.78(-3)$ & $3.42(-3)$ & $-28.6$ & $1.01$ & $9.22(-1)$ & $-9.1$ \\
19 & $3.85(-3)$ & $2.76(-3)$ & $-28.4$ & $1.00$ & $9.14(-1)$ & $-8.9$ \\
20 & $3.10(-3)$ & $2.23(-3)$ & $-28.2$ & $9.91(-1)$ & $9.06(-1)$ & $-8.7$ \\
21 & $2.50(-3)$ & $1.80(-3)$ & $-28.0$ & $9.80(-1)$ & $8.97(-1)$ & $-8.5$ \\
22 & $2.02(-3)$ & $1.46(-3)$ & $-27.8$ & $9.69(-1)$ & $8.89(-1)$ & $-8.3$ \\
23 & $1.64(-3)$ & $1.18(-3)$ & $-27.6$ & $9.59(-1)$ & $8.82(-1)$ & $-8.1$ \\
24 & $1.33(-3)$ & $9.62(-4)$ & $-27.5$ & $9.49(-1)$ & $8.74(-1)$ & $-7.9$ \\
25 & $1.08(-3)$ & $7.84(-4)$ & $-27.3$ & $9.40(-1)$ & $8.67(-1)$ & $-7.7$ \\
26 & $8.79(-4)$ & $6.40(-4)$ & $-27.2$ & $9.31(-1)$ & $8.61(-1)$ & $-7.5$ \\
27 & $7.18(-4)$ & $5.24(-4)$ & $-27.0$ & $9.23(-1)$ & $8.55(-1)$ & $-7.4$ \\
28 & $5.87(-4)$ & $4.29(-4)$ & $-26.9$ & $9.16(-1)$ & $8.50(-1)$ & $-7.2$ \\
29 & $4.80(-4)$ & $3.51(-4)$ & $-26.8$ & $9.10(-1)$ & $8.46(-1)$ & $-7.1$ \\
30 & $3.91(-4)$ & $2.87(-4)$ & $-26.6$ & $9.05(-1)$ & $8.42(-1)$ & $-7.0$ \\
31 & $3.16(-4)$ & $2.32(-4)$ & $-26.5$ & $9.00(-1)$ & $8.38(-1)$ & $-6.9$ \\
32 & $2.52(-4)$ & $1.85(-4)$ & $-26.4$ & $8.96(-1)$ & $8.35(-1)$ & $-6.8$ \\
33 & $1.93(-4)$ & $1.42(-4)$ & $-26.3$ & $8.93(-1)$ & $8.33(-1)$ & $-6.7$ \\
34 & $1.36(-4)$ & $1.00(-4)$ & $-26.3$ & $8.90(-1)$ & $8.31(-1)$ & $-6.6$ \\
35 & $7.41(-5)$ & $5.46(-5)$ & $-26.2$ & $8.88(-1)$ & $8.30(-1)$ & $-6.6$ \\
36 & $2.05(-5)$ & $1.51(-5)$ & $-26.2$ & $8.87(-1)$ & $8.29(-1)$ & $-6.6$ \\
\hline
\end{tabular}
\end{table*}

\begin{table*}[htbp]
\centering
\caption{Same as Table~\ref{tab:ratios_B}, but for vibrational transitions $X(v_l=0)\rightarrow C(v_u)$.}
\label{tab:ratios_C}
\begin{tabular}{ccccccc}
\hline
$v_u$ & \multicolumn{2}{c}{$\zeta^{\text{exc}}_{{\rm H}_2}/\zeta^{\text{ion}}_{{\rm H}_2}$} & Rel. diff. (\%) & \multicolumn{2}{c}{$\zeta^{\rm exc}_{{\rm sec},{\rm H}_2}/\zeta^{\rm exc}_{p,{\rm H}_2}$} & Rel. diff. (\%) \\
\cline{2-3}\cline{5-6}
& exact & SS approx. & & exact & SS approx. & \\
\hline\hline
0 & $5.33(-2)$ & $3.75(-2)$ & $-29.7$ & $1.06$ & $9.67(-1)$ & $-9.1$ \\
1 & $7.88(-2)$ & $5.58(-2)$ & $-29.3$ & $1.03$ & $9.43(-1)$ & $-8.8$ \\
2 & $7.31(-2)$ & $5.20(-2)$ & $-28.8$ & $1.01$ & $9.27(-1)$ & $-8.5$ \\
3 & $5.57(-2)$ & $3.98(-2)$ & $-28.5$ & $9.92(-1)$ & $9.10(-1)$ & $-8.2$ \\
4 & $3.83(-2)$ & $2.76(-2)$ & $-28.1$ & $9.71(-1)$ & $8.94(-1)$ & $-8.0$ \\
5 & $2.51(-2)$ & $1.81(-2)$ & $-27.8$ & $9.54(-1)$ & $8.80(-1)$ & $-7.7$ \\
6 & $1.60(-2)$ & $1.16(-2)$ & $-27.5$ & $9.39(-1)$ & $8.69(-1)$ & $-7.5$ \\
7 & $1.01(-2)$ & $7.36(-3)$ & $-27.2$ & $9.27(-1)$ & $8.59(-1)$ & $-7.3$ \\
8 & $6.40(-3)$ & $4.68(-3)$ & $-26.9$ & $9.15(-1)$ & $8.50(-1)$ & $-7.1$ \\
9 & $4.07(-3)$ & $2.99(-3)$ & $-26.7$ & $9.03(-1)$ & $8.41(-1)$ & $-6.9$ \\
10 & $2.61(-3)$ & $1.92(-3)$ & $-26.5$ & $8.93(-1)$ & $8.33(-1)$ & $-6.7$ \\
11 & $1.65(-3)$ & $1.22(-3)$ & $-26.3$ & $8.83(-1)$ & $8.25(-1)$ & $-6.6$ \\
12 & $1.00(-3)$ & $7.42(-4)$ & $-26.2$ & $8.76(-1)$ & $8.20(-1)$ & $-6.4$ \\
13 & $4.93(-4)$ & $3.65(-4)$ & $-26.1$ & $8.71(-1)$ & $8.16(-1)$ & $-6.4$ \\
\hline
\end{tabular}
\end{table*}

\subsection{Spectrum of secondary electrons}
\label{sect:jsec}

In this Section we show that the ratio $\zeta^{\rm exc}_{\ce{H2}}/\zeta^{\rm ion}_{\ce{H2}}$ is independent on the 
assumed cosmic-ray proton spectrum and \ce{H2} column density at which it is computed. We first prove this through an analytical procedure, recalling a previous approximation for calculating the secondary electron spectrum (Sect.~\ref{sect:OSA}), and then introducing a new, more accurate approximation (Sect.~\ref{sect:ESSA}).
In the following we only consider secondary electrons from proton-impact \ce{H2} ionisation. We note however that the excitation by 
secondary electrons from electron-impact ionisation
is not negligible below \ce{H2} column densities of about 10$^{21}$~cm$^{-2}$, on the basis of the primary spectra measured by the Voyager probes.

\subsubsection{On-the-spot approximation}
\label{sect:OSA}

The flux of secondary electrons can be obtained in the on-the-spot (OS) 
approximation\cite{Ivlev+2015,PadovaniGaches2024} as
\be
j_{\rm sec,\ce{H2}}(\varepsilon) \approx \frac{\varepsilon}{L_{e,{\rm H}_2}(\varepsilon)}\int_{I^{\rm ion}_{{\rm H}_2}+\varepsilon}^\infty j_p(E)\frac{\partial \sigma^{\rm ion}_{p,{\rm H}_2}}{\partial\varepsilon}(E,\varepsilon)\,\ud E\,,
\label{eq:osa}
\ee
where $L_{e,{\rm H}_2}(\varepsilon)$ is the energy loss function of electrons in \ce{H2} and $\partial\sigma^{\rm ion}_{p,{\rm H}_2}(E,\varepsilon)/\partial\varepsilon$ 
is the single differential cross section for ionisation of H$_2$ by proton impact.
We assume that the differential cross section can be approximately factorised as
\be
\frac{\partial\sigma^{\rm ion}_{p,{\rm H}_2}(E,\varepsilon)}{\partial\varepsilon}\approx\sigma^{\rm ion}_{p,{\rm H}_2}(E)\varphi_{\ce{H2}}(\varepsilon)\,,
\label{eq:phi}
\ee
which is valid for $E\gtrsim1$~MeV\cite{Rudd1988} 
(note that the largest contribution to H$_2$ ionisation comes from protons in this specific energy range\cite{Padovani+2022}).
Then, Eq.~(\ref{eq:osa}) becomes
\begin{eqnarray}\label{eq:jsec_osa}
j_{\rm sec,\ce{H2}}(\varepsilon) &\approx& \frac{\varepsilon\varphi_{\ce{H2}}(\varepsilon)}{L_{e,{\ce{H2}}}(\varepsilon)}\int_{I^{\rm ion}_{{\rm H}_2}+\varepsilon}^\infty j_p(E)\sigma^{\rm ion}_{p,{\rm H}_2}(E)\,\ud E \\\nonumber
&&=\frac{\zeta^{\rm ion}_{p,{\rm H}_2}}{4\pi L_{e,\ce{H2}}(\varepsilon)}\varepsilon\varphi_{\ce{H2}}(\varepsilon)  \,,
\end{eqnarray}
where we have set $\zeta^{\rm ion}_{p,\ce{H2}}=\zeta^{\rm ion}_{p,\ce{H2}}(\varepsilon=0)$ 
since the function $\varphi_{\ce{H2}}(\varepsilon)$ drops by orders of magnitude above 100~eV (see Fig.~\ref{fig:phi_Phi_epsilonphi}). 
In fact, the typical energy of secondary electrons, $\varepsilon\approx30-50$~eV, is much smaller than the energy of protons that dominate $\zeta^{\rm ion}_{p,{\rm H}_2}$\cite{Padovani+2022}. 
Thus, adding $\varepsilon$ to $I^{\rm ion}_{{\rm H}_2}$ 
increases the lower limit by a negligible amount relative to the energies that contribute the most to the integral.

\subsubsection{Steady-state approximation}
\label{sect:ESSA}

Here we introduce a more general approach starting from the transport equation for secondary electrons in steady state,
\be
\frac{\ud}{\ud\varepsilon}\left[\left(\frac{\ud\varepsilon}{\ud t}\right)_{\rm loss}\mathcal{N}_{\rm sec,\ce{H2}}(\varepsilon)\right] = Q(\varepsilon)\,,
\label{eq:steadystate}
\ee
where $(\ud\varepsilon/\ud t)_{\rm loss}=-n({\rm H}_2)v(\varepsilon)L_{e,{\rm H}_2}(\varepsilon)$ is the energy lost per unit time, $n({\rm H}_2)$ is the H$_2$ volume density,
$v(\varepsilon)$ is the electron velocity, 
$\mathcal{N}_{\rm sec,\ce{H2}}(\varepsilon)$ is the number of secondary electrons per unit volume and energy, 
and
\be
Q(\varepsilon) = 4\pi n({\rm H}_2) \int_{I_{{\rm H}_2}^{\rm ion}+\varepsilon}^\infty j_p(E) \frac{\partial\sigma^{\rm ion}_{p,{\rm H}_2}}{\partial\varepsilon}(E,\varepsilon)\,\ud E
\label{eq:Q}
\ee
is the rate of injection of electrons per unit volume and energy.
Substituting Eq.~(\ref{eq:Q}) in Eq.~(\ref{eq:steadystate}), one can solve for $\mathcal{N}_{\rm sec,\ce{H2}}(\varepsilon)$. Then, since $\mathcal{N}_{\rm sec,\ce{H2}}(\varepsilon)=4\pi j_{\rm sec,\ce{H2}}(\varepsilon)/v(\varepsilon)$ and again assuming $\zeta^{\rm ion}_{p,{\rm H}_2}(\varepsilon)=\zeta^{\rm ion}_{p,{\rm H}_2}$, the secondary electron spectrum 
obtained from the steady-state (SS) approximation is
\be
j_{\rm sec,\ce{H2}}(\varepsilon) \approx \frac{\zeta^{\rm ion}_{p,{\rm H}_2}}{4\pi L_{e,{\rm H}_2}(\varepsilon)}\Phi_{\ce{H2}}(\varepsilon)\,,
\label{eq:jsec_essa}
\ee
where
\be
\Phi_{\ce{H2}}(\varepsilon) = \int_\varepsilon^\infty \varphi_{\ce{H2}}(\varepsilon)\ud\varepsilon\,.
\label{eq:Phi}
\ee

\begin{figure}
\includegraphics[width=0.5\textwidth]{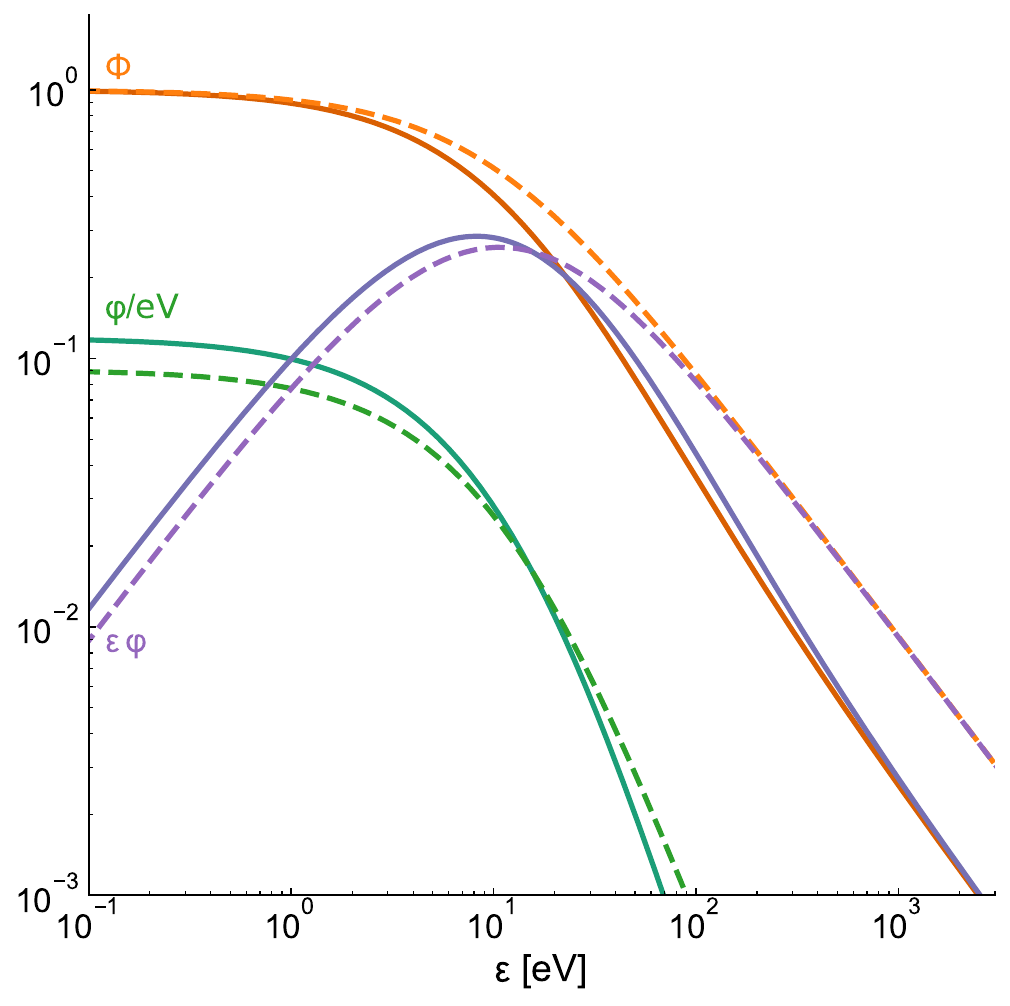}
\caption{Auxiliary functions for the secondary electron spectrum:
$\varphi_{\ce{H2}}(\varepsilon)$ (Eq.~\ref{eq:phi}), 
$\varepsilon\varphi_{\ce{H2}}(\varepsilon)$ (Eq.~\ref{eq:jsec_osa}), 
and $\Phi_{\ce{H2}}(\varepsilon)$ (Eqs.~\ref{eq:jsec_essa},~\ref{eq:Phi}). The dashed curves show the same quantities according to the BEA 
(Eqs.~\ref{eq:phibeb},~\ref{eq:PHIbeb}).}
\label{fig:phi_Phi_epsilonphi}
\end{figure}

\begin{figure}
\includegraphics[width=0.5\textwidth]{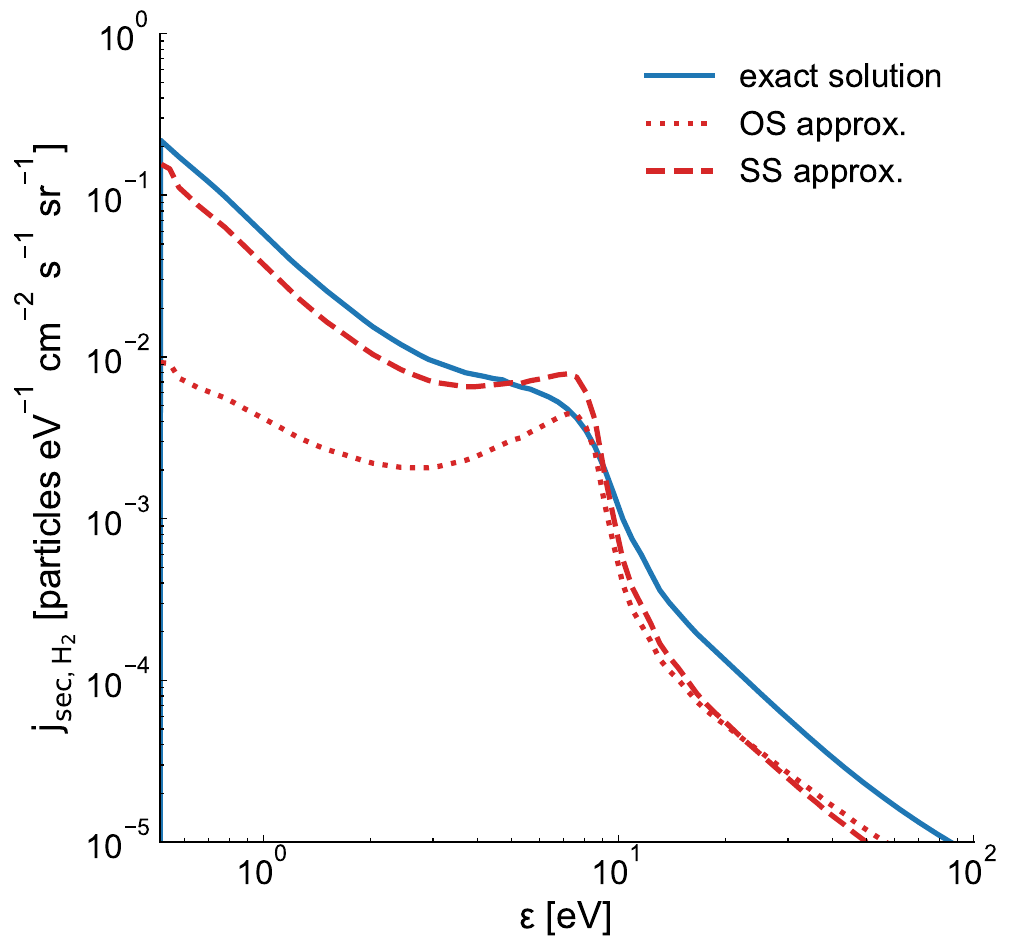}
\caption{Flux of secondary electrons computed exactly by solving the balance equation\cite{Ivlev+2021} 
(solid blue line),
using the on-the-spot (OS) approximation (dotted red line, Eq.~\ref{eq:jsec_osa}), 
and the steady-state (SS) approximation (dashed red line, Eq.~\ref{eq:jsec_essa}).
The input cosmic-ray proton spectrum is the model $\mathscr{L}$ at 
$N(\ce{H2})=10^{22}$~cm$^{-2}$.}
\label{fig:jsec_exact_ESA_OSSA}
\end{figure}

The OS (Eq.~\ref{eq:jsec_osa}) and SS (Eq.~\ref{eq:jsec_essa}) approximations differ by the last factor.
As shown in Fig.~\ref{fig:phi_Phi_epsilonphi}, $\Phi_{\ce{H2}}(\varepsilon)\simeq\varepsilon\varphi_{\ce{H2}}(\varepsilon )$ above the \ce{H2} ionisation threshold, 
while at lower energies the difference becomes increasingly larger.
Consequently, the OS approximation underestimates the spectrum of secondary electrons at energies that are important for excitation processes,
while the SS approximation accurately reproduces the exact solution obtained by solving the balance equation\cite{Ivlev+2021} (see Fig.~\ref{fig:jsec_exact_ESA_OSSA}). 

For the calculation of the functions $\varphi_{\ce{H2}}(\varepsilon)$ and $\Phi_{\ce{H2}}(\varepsilon)$ we have adopted the empirical differential cross section based on the so-called ``molecular promotion model''\cite{Rudd1988}. The classical ``binary encounter approximation''\cite{Williams1927} (BEA) gives an exact factorisation of the differential cross section in the form of Eq. (\ref{eq:phi}), valid for $E\gg (m_p/4m_e)I^{\rm ion}_{{\rm H}_2}\approx 7$~keV, with $\sigma^{\rm ion, BEA}_{p,\ce{H2}}(E)\propto E^{-1}$ and 
\be
\varphi^{\rm BEA}_{p,\ce{H2}}(\varepsilon)=\frac{I^{\rm ion}_{\ce{H2}}(7I^{\rm ion}_{\ce{H2}}+3\varepsilon)}{5(I^{\rm ion}_{\ce{H2}}+\varepsilon)^3}\,.
\label{eq:phibeb}
\ee
This gives
\be
\Phi^{\rm BEA}_{\ce{H2}}(\varepsilon)=\frac{I^{\rm ion}_{\ce{H2}}(5I^{\rm ion}_{\ce{H2}}+3\varepsilon)}{5(I^{\rm ion}_{\ce{H2}}+\varepsilon)^2}.
\label{eq:PHIbeb}
\ee
The BEA values of $\varphi(\varepsilon)$ and $\Phi(\varepsilon)$ are shown in Fig.~\ref{fig:phi_Phi_epsilonphi} by dashed curves. Clearly, the classical BEA fails to reproduce the high-energy behaviour of the ionisation cross section, which is important to determine the spectrum of secondary electrons.

We note that both the OS and SS approximations are proportional to $\zeta^{\rm ion}_{p,{\rm H}_2}$ as expected on the basis of the universal degradation spectrum theory\cite{SpencerFano1954},
according to which the steady degradation spectrum is governed by an integral equation whose solution is a universal yield spectrum,
and the fractions of energy lost in ionisation, excitation, and heating become nearly independent of the injection spectrum. 

Substituting Eq.~(\ref{eq:jsec_essa}) in Eqs.~(\ref{eq:zexc}) and (\ref{eq:zion}), one gets
\be
\zeta^{\rm exc}_{{\rm H}_2} = \zeta^{\rm ion}_{p,{\rm H}_2} (\mk^{\rm exc}_{p,{\rm H}_2} + \mk^{\rm exc}_{{\rm sec},{\rm H}_2})\nonumber
\ee
and
\be
\zeta^{\rm ion}_{{\rm H}_2} = \zeta^{\rm ion}_{p,{\rm H}_2} (1 + \mk^{\rm ion}_{{\rm sec},{\rm H}_2})\,,\nonumber
\ee
where $\mk^{\rm exc}_{{\rm sec},{\rm H}_2}$ and $\mk^{\rm ion}_{{\rm sec},{\rm H}_2}$ are constants equal to
\be
\mk^{\rm exc}_{{\rm sec},{\rm H}_2} = \int_{I^{\rm exc}_{{\rm H}_2}} \frac{\Phi_{\ce{H2}}(\varepsilon)}{L_{e,\ce{H2}}(\varepsilon)}\sigma^{\rm exc}_{e,{\rm H}_2}(\varepsilon)\,\ud\varepsilon\,,\nonumber
\label{eq:Cexce}
\ee
and
\be
\mk^{\rm ion}_{{\rm sec},{\rm H}_2} = \int_{I^{\rm ion}_{{\rm H}_2}} \frac{\Phi_{\ce{H2}}(\varepsilon)}{L_{e,{\rm H}_2}(\varepsilon)}\sigma^{\rm ion}_{e,{\rm H}_2}(\varepsilon)\,\ud\varepsilon\,,\nonumber
\label{eq:Cione}
\ee
respectively. Also $\mk^{\rm exc}_{p,{\rm H}_2} = \sigma^{\rm exc}_{p,{\rm H}_2}(E)/\sigma^{\rm ion}_{p,{\rm H}_2}(E)$ is nearly constant 
for $E\gtrsim1$~MeV 
as per the Bethe-Born approximation.
This shows that the ratio
\be
\frac{\zeta^{\rm exc}_{{\rm H}_2}}{\zeta^{\rm ion}_{{\rm H}_2}} \approx \frac{\mk^{\rm exc}_{p,{\rm H}_2} + \mk^{\rm exc}_{{\rm sec},{\rm H}_2}}{1+\mk^{\rm ion}_{{\rm sec},{\rm H}_2}}\nonumber
\ee
is independent of the particular shape of the input proton spectrum, $j_p$.
Similarly, $\zeta^{\rm exc}_{{\rm sec},{\rm H}_2}/\zeta^{\rm exc}_{p,{\rm H}_2}\approx \mk^{\rm exc}_{{\rm sec},{\rm H}_2}/\mk^{\rm exc}_{p,{\rm H}_2}$ is a constant.

The constant $\mk^{\rm ion}_{{\rm sec},{\rm H}_2}$ is exactly the ratio $\zeta^{\rm ion}_{{\rm sec},\ce{H2}}/\zeta^{\rm ion}_{p,{\rm H}_2}$. 
We find $\mk^{\rm ion}_{{\rm sec},{\rm H}_2}=0.73$, in excellent agreement with 
previous estimates\cite{DalgarnoGriffing1958,Ivlev+2021}.
Tables~\ref{tab:ratios_B} and~\ref{tab:ratios_C} also show the values of $\zeta^{\rm exc}_{{\rm H}_2}/\zeta^{\rm ion}_{{\rm H}_2}$ and $\zeta^{\rm exc}_{{\rm sec},{\rm H}_2}/\zeta^{\rm exc}_{p,{\rm H}_2}$ using the
SS approximation, which is found to be accurate to better than 30\% 
for $\zeta^{\rm exc}_{{\rm H}_2}/\zeta^{\rm ion}_{{\rm H}_2}$ and better than about 15\% for $\zeta^{\rm exc}_{{\rm sec},{\rm H}_2}/\zeta^{\rm exc}_{p,{\rm H}_2}$
for both sets of vibrational transitions.

\subsubsection{Proton-impact ionisation of atomic hydrogen}
\label{sect:hion}

Finally, we use the SS approximation to compute ratios between the ionisation rate of atomic and molecular hydrogen by cosmic-ray protons. 
In the literature, these two quantities are usually referred to as $\zeta_1$ and $\zeta_2$, respectively. 

\begin{figure}
\includegraphics[width=0.5\textwidth]{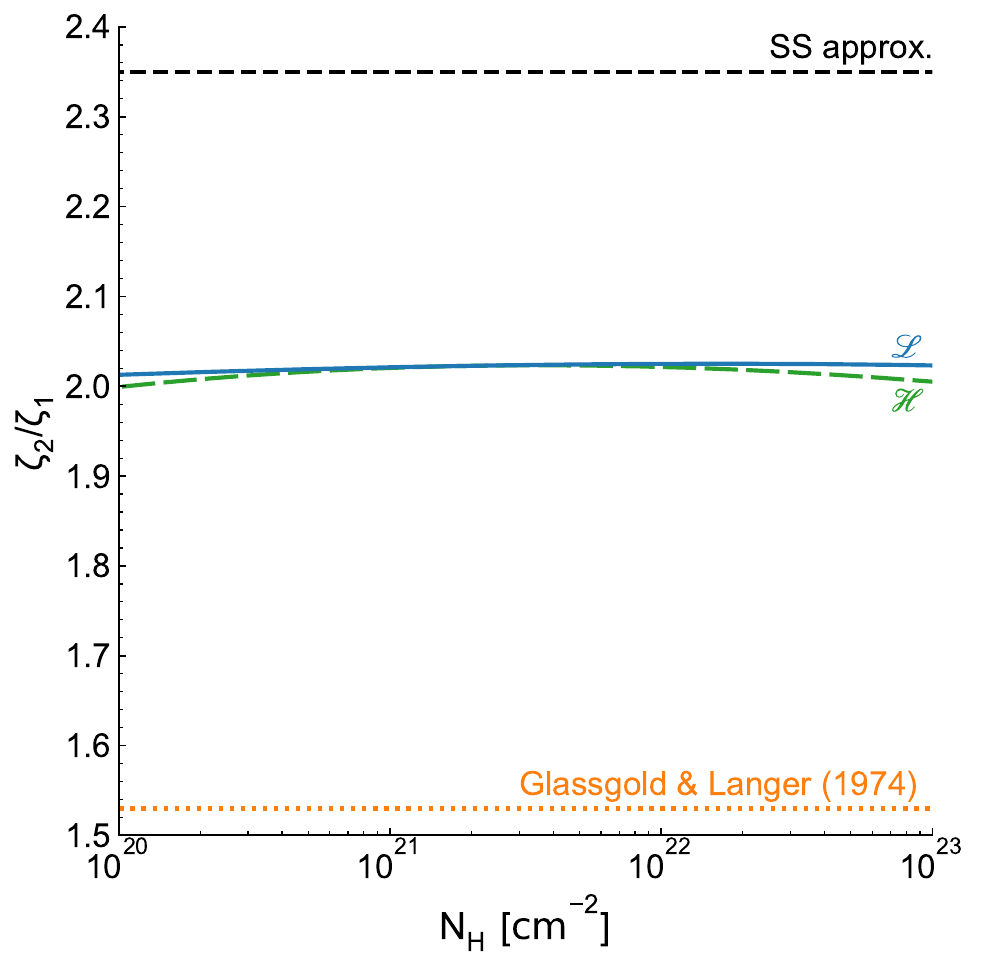}
\caption{Ratio between the cosmic-ray ionisation rate of \ce{H2} (primary plus secondary contributions, $\zeta_2$)
and of \ce{H} (primary plus secondary contributions, $\zeta_1$) as a function of the total hydrogen
column density, $N_{\ce{H}}$, obtained from the exact solution
for the model $\mathscr{L}$ and $\mathscr{H}$ (solid blue and long-dashed green line, respectively) and
from the SS approximation
(short-dashed black line, Eq.~\ref{eq:jsec_essa}) 
compared to the canonical value\cite{GlassgoldLanger1974} 
(dotted orange line).}
\label{fig:z2z1_comparison_GL74}
\end{figure}

Let us define
\be
\zeta^{\rm ion}_{p,\ce{H}} = 4\pi \int_{I^{\rm ion}_{\ce{H}}} j_p(E) \sigma^{\rm ion}_{p,\ce{H}}(E) \ud E\,,\nonumber
\ee
where $\sigma^{\rm ion}_{p,\ce{H}}(E)$ is the proton-impact ionisation cross section of H, and $I^{\rm ion}_{\ce{H}}=13.6$~eV is the ionisation energy 
threshold of \ce{H} ($\zeta^{\rm ion}_{p,\ce{H}}$ is usually referred to as $\zeta_p$ in the literature). 
Applying the same factorisation as in Eq.~(\ref{eq:phi}) to 
the single differential ionisation cross section of atomic hydrogen above about 1~MeV, we obtain
\be
\frac{\zeta_2}{\zeta_1} = \frac{\zeta^{\rm ion}_{p,\ce{H2}} (1 + \mk^{\rm ion}_{{\rm sec},\ce{H2}})}{\zeta^{\rm ion}_{p,\ce{H}} (1 + \mk^{\rm ion}_{{\rm sec},\ce{H}})}\,,
\label{eq:z2z1}
\ee
where 
\be
\mk^{\rm ion}_{{\rm sec},\ce{H}} = \int_{I_{\rm ion,\ce{H}}} \frac{\Phi_{\ce{H}}(\varepsilon)}{L_{e,\ce{H}}(\varepsilon)}
\sigma^{\rm ion}_{e,\ce{H}}(\varepsilon)\ud\varepsilon\,,\nonumber
\ee
and
\be
\Phi_{\ce{H}}(\varepsilon) = \int_{\varepsilon}^\infty \varphi_{\ce{H}}(\varepsilon)\ud\varepsilon\,.\nonumber
\ee
It turns out that $1+\mk^{\rm ion}_{{\rm sec},\ce{H2}} \approx 1+\mk^{\rm ion}_{{\rm sec},\ce{H}}$ within less than 20\%, 
and $\zeta^{\rm ion}_{p,\ce{H2}}\approx 2\zeta^{\rm ion}_{p,\ce{H}}$, since $\sigma^{\rm ion}_{p,\ce{H2}}\approx 2\sigma^{\rm ion}_{p,\ce{H}}$ at
$E\gtrsim1$~MeV as per the Bethe-Born approximation.
Figure~\ref{fig:z2z1_comparison_GL74} shows the ratio $\zeta_2/\zeta_1$ as a function of the total (atomic plus molecular) 
hydrogen column density, $N_{\ce{H}} = N(\ce{H})+2N(\ce{H2})$.
The canonical value\cite{GlassgoldLanger1974} of $\zeta_2/\zeta_1=1.53$ is compared to
our SS approximation ($\zeta_2/\zeta_1=2.35$, Eq.~\ref{eq:z2z1}), and to the exact solution obtained by solving the balance equation\cite{Ivlev+2021} for the spectrum of secondary 
electrons produced by proton-impact ionisation of atomic hydrogen ($\zeta_2/\zeta_1\approx 2$). The latter value
is practically independent of the cosmic-ray spectrum and the total hydrogen column density.

\section{Discussion and Conclusions}
\label{sect:conclusions}

Cosmic rays are recognised as fundamental drivers of the complex chemical and dynamic processes governing star-forming regions. 
The primary ionisation of molecular hydrogen generates a strong flux of secondary 
electrons at very low energies ($<100$~eV). These electrons have been studied primarily 
for their ability to excite the rovibrational and electronic states of \ce{H2}, 
influencing the observed spectral signatures and driving chemical pathways. 
Electron-impact excitation cross sections of \ce{H2} have been 
the subject of extensive theoretical and experimental investigations, providing a solid framework for understanding energy transfer processes in the interstellar medium. 
Traditionally, electrons have been the primary focus due to their lower mass and greater mobility than nuclei, which allows more frequent collisions 
than nuclei
with hydrogen molecules. 
The commonly adopted approximation for estimating the proton-impact excitation was to consider 
their cross sections to be the same as those of electrons of the same velocity.
In this article, we have used a set of cross sections specifically developed to enable the study of collisions between heavy particles and diatomic molecules. In particular, we focused on the excitation of the electronic states $B$ and $C$ of \ce{H2} by proton impact.
We calculated the \ce{H2} excitation rates with this new set of cross sections, revealing a significant 
contribution from protons. Specifically, we discovered that the contribution to the total excitation rate of \ce{H2} by protons is comparable to that of electrons and, for some vibrational transitions, even higher. 

This result has potentially 
significant implications in a number of astrophysical contexts. For example, if this same result were confirmed for transitions occurring in the ground state of \ce{H2}, we could predict an increase in the $\zeta^{\rm exc}_{\ce{H2}}/\zeta^{\rm ion}_{\ce{H2}}$ ratio. The latter provides an essential link between the observed intensity of a NIR spectral line and $\zeta^{\rm ion}_{\ce{H2}}$. Consequently, an increase in $\zeta^{\rm exc}_{\ce{H2}}$ would require a proportional decrease in $\zeta^{\rm ion}_{\ce{H2}}$ to maintain consistency with the observed line intensity. This re-evaluation of $\zeta^{\rm ion}_{\ce{H2}}$ could lead to significant revisions in our understanding of ionisation mechanisms and energy balance in environments where \ce{H2} excitation 
by cosmic rays is important.
The availability of a new database of rotationally-resolved excitation cross sections will be also crucial to advance our understanding of how proton excitation influences the intensity of the ultraviolet \ce{H2} emission as well as the corresponding modifications in photodissociation and photoionisation rates of atoms and molecules of astrophysical interest.

In this article, we also demonstrate how the ratio $\zeta^{\rm exc}_{\ce{H2}}/\zeta^{\rm ion}_{\ce{H2}}$ is independent of the choice of Galactic cosmic-ray proton spectrum and molecular hydrogen column density at 
which it propagates. Through this proof, we are also able to provide an approximate analytical solution for the secondary electron spectrum that avoids solving the integral-differential balance equation and can therefore be easily implemented in numerical codes.

\begin{acknowledgements}

The authors wish to thank the referees for their
careful reading of the manuscript and insightful comments.
M.P. acknowledges the INAF grant 2023 MERCATOR (``MultiwavelEngth signatuRes of Cosmic rAys in sTar-fOrming Regions'') and 
the INAF grant 2024 ENERGIA (``ExploriNg low-Energy cosmic Rays throuGh theoretical InvestigAtions at INAF'');
D.G. acknowledges the INAF grant 2023 PACIFISM (``PArtiCles, Ionization and Fields in the InterStellar Medium'').
The MCCC research was supported by the Australian Government through the Australian Research Council’s 
Discovery Projects funding scheme (project DP240101184). 
L.H.S. is the recipient of an Australian Research Council Discovery Early Career Researcher Award 
(project number DE240100176) funded by the Australian Government. 
M.C.Z. would like to specifically acknowledge Los Alamos National Laboratory’s ASC PEM Atomic Physics Project and Laboratory
Directed Research and Development program Project No. 20240391ER. 
LANL is operated by Triad National Security, LLC, for the National Nuclear Security 
Administration of the U.S. Department of Energy under Contract No. 89233218NCA000001. 
HPC resources were provided by the Pawsey Supercomputing Centre with funding from the Australian Government and the Government of Western Australia, and the Texas Advanced Computing Center at the University of Texas at Austin.

\end{acknowledgements}

\newpage

\bibliographystyle{abbrvnat}
\bibliography{mybibliography-bibdesk}

\end{document}